\documentclass[useAMS,usenatbib,onecolumn]{mn2e}

\usepackage{graphicx}
\def\om{\eps}\def\ome{\om_e}\def\cte{\cos\theta_e}
\def\ctqe{\cos^2\theta_e}\def\stqe{\sin^2\theta_e}
\def\eps{\epsilon}\def\vphi{\varphi}
\def\=={\ =\ }\def\--{\ -\ }\def\++{\ +\ }\def\su#1#2{\displaystyle{{#1}\over{#2}}}
\def\ct{\,\cos\theta}\def\st{\,\sin\theta}\def\ctq{\,\cos^2\theta}\def\stq{\,\sin^2\theta}
\let\gsim\ga
\def\Tn#1{\,T^{n,\ell}_{#1}}\def\Gn#1{\,\Gamma^{n}_{#1}}
\def\RN#1{{\cal U\,}_{#1}}\def\bk{{\beta^{(n)}_k}}
\def\Ome{\Omega}\def\V#1{{\cal M}_{#1}\,}
\def\Mat#1#2#3#4{\left(\matrix{#1 & #2 \cr #3 & #4 \cr }\right)}
\def\PMQ{G\,}\def\PPQ{H\,}
\def\ERF{ERF}\def\ERFS{ERF$^\star$}\def\ERFSS{ERF$^{\star\star}$}

\title[X-ray spectra from magnetar candidates. II]{
X-ray spectra from magnetar candidates.
II. Resonant cross sections for electron-photon scattering in the
relativistic regime}
\author[L. Nobili, R. Turolla and S. Zane]{L. Nobili$^{1}$\thanks{E-mail: nobili@pd.infn.it (LN);
turolla@pd.infn.it (RT); sz@mssl.ucl.ac.uk (SZ)}, R. Turolla$^{1, 2}$ and S. Zane$^{2}$\\ $^{1}$Department of
Physics, University of Padova, via Marzolo 8, 35131 Padova, Italy\\ $^{2}$Mullard Space Science Laboratory,
University College London, Holmbury St. Mary, Dorking, Surrey, RH5 6NT, UK}

\begin{document}

\date{}

\pagerange{\pageref{firstpage}--\pageref{lastpage}}\pubyear{2008}
\maketitle \label{firstpage}
\begin{abstract}

Recent models of spectral formation in magnetars called renewed attention on electron-photon scattering in the
presence of ultra-strong magnetic fields. Investigations presented so far mainly focussed on mildly relativistic
particles and magnetic scattering was treated in the non-relativistic (Thomson) limit. This allows for
consistent spectral calculations up to a few tens of keVs, but becomes inadequate in modelling the hard tails
($\la 200$ keV) detected by {\em INTEGRAL} from magnetar sources.
In this paper, the second in a series devoted to model the X-/soft
$\gamma$-ray persistent spectrum of magnetar candidates, we present
explicit, relatively simple expressions for the magnetic Compton
cross-section at resonance which account for Landau-Raman scattering up to
the second Landau level. No assumption is made on the magnetic field
strength.  We find that sensible departures from the Thomson 
regime can be already 
present at $B\sim 5\times 10^{12}$~G. The form of the magnetic cross section we derived can be easily implemented
in Monte Carlo transfer codes and 
a direct application to magnetar spectral calculations will be presented in a forthcoming study.
\end{abstract}

\section{Introduction}\label{intro}

The recent discovery with the {\em INTEGRAL} satellite of hard X-ray tails
in the (persistent) spectra of the magnetar candidates (the anomalous
X-ray pulsars, AXPs, and the soft $\gamma$-repeaters, SGRs; e.g.
\citealt{sandrorev}) provides evidence that a sizeable fraction (up to
$\sim 50\%$) of the energy output of these sources is emitted above $\sim
20$ keV. Up to now, high-energy emission has been detected in two SGRs,
1806-20 and 1900+14
\citep{sandro05, diego06}, 
and three AXPs, 4U 0142+614, 1RXS J170849-4009 and 1E
1841-045 \cite[][see also for an updated summary of {\em INTEGRAL}
observations \citealt{diego08}]{kui04, kui06}
\footnote{Very recently \cite{Leyder08} reported the
{\em INTEGRAL} detection of the AXP 1E~1048.1-5937, but no spectral
information are available yet.}. This result come somehow unexpected, since the
spectra of SGRs/AXPs in the soft X-ray range ($\sim 0.1$--10 keV) are well
described by a two component model, a blackbody at $kT\sim 0.2$--0.6 keV,
and a rather steep power-law with photon index $\Gamma_{soft}\sim 1.5$--4.
{\em INTEGRAL} observations have shown that in SGRs the power-law tail
extends unbroken (or steepens) in the $\sim 20$--200 keV range,
$\Gamma_{hard}\ga\Gamma_{soft}\sim 1.5$. In AXPs, which have steeper
spectra in the soft X-ray band, a spectral upturn appears, i.e. the
high-energy power-law is harder than the soft one, $\Gamma_{hard}\sim 1$
while $\Gamma_{soft}\sim 3$--4 \cite[see e.g.][for a joint spectral
analysis of {\em XMM-Newton} and {\em INTEGRAL} data]{nanda08}.

Within the magnetar scenario, the persistent 0.1--10 keV emission of SGRs and AXPs has been successfully
interpreted in terms of the twisted magnetosphere model \citep{tlk02}. In an ultra-magnetized neutron star the
huge toroidal field stored in the interior produces a deformation of the crust. The displacement of the
footprints of the external (initially dipolar) field drives the growth of a toroidal component which, in turn,
requires supporting currents. Charges flowing in the magnetosphere provide a large optical depth to resonant
cyclotron scattering (RCS) and repeated scatterings of thermal photons emitted by the star surface may then lead
to the formation of a power-law tail. The original picture by \cite{tlk02} has been further
explored by \cite{lg06}, \cite{ft07} and \cite{papI}.
Recently \cite{nanda08} presented a systematic application
of the 1D, analytical RCS model of \cite{lg06} to a large sample of magnetars X-ray spectra finding a good
agreement with data in the 0.1--10 keV range. As shown in paper I, more sophisticated 3D Monte Carlo calculations
of RCS spectra successfully reproduce soft X-ray observations.

Despite the twisted magnetosphere scenario appears quite promising in explaining the magnetars soft X-ray
emission, if and how it can account also for the hard tails detected with {\em INTEGRAL} has not been
unambiguously shown as yet. \cite{tb05} suggested that the hard X-rays may be produced either by thermal
bremsstrahlung in the surface layers heated by returning currents, or by synchrotron emission from pairs created
higher up ($\approx 100$ km) in the magnetosphere. 
\cite{bh07,bh08} have recently proposed a further possibility,
according to which the soft $\gamma$-rays may originate from resonant up-scattering of seed photons on a
population of highly relativistic electrons. Previous investigations (\citealt{lg06}; \citealt{ft07}; paper I)
mainly focussed on mildly relativistic particles and magnetic scattering was treated in the non-relativistic
(Thomson) limit. This is perfectly adequate in assessing the spectral properties up to energies $\ll
m_ec^2/\gamma$ (here $\gamma$ is the typical electron Lorentz factor) since i) the energy of primary photons is
low enough ($\approx 1$ keV) to make resonant scattering onto electrons possible only where the magnetic field
has dropped well below the QED critical field, ii) up-scattering onto electrons with $\gamma\approx 1$ hardly
boosts the photon energy in the hundred keV range, so electron recoil is not important. However, some photons do
actually gain quite a large energy (because they scatter many times on the most energetic electrons) and fill a
tail at energies $\ga 50$ keV. We caveat that, despite in previous works
spectra have been computed up to the MeV range, they become untrustworthy above some tens of keV and
can not be used to assess the spectral features that can arise due to electron recoil effects (i.e. a
high-energy spectral break). Proper investigation of the latter demands a complete QED treatment of magnetic
Compton scattering. This is mandatory if highly relativistic particles are considered because a photon can be
boosted to quite large energies in a single scattering and, if it propagates towards the star, it may scatter
again where the field is above the QED limit.

Monte Carlo numerical codes for radiation transport in a magnetized,
scattering medium, as the one we presented in paper I, make an excellent tool to
investigate in detail the properties of RCS in the case in which energetic electrons
are present in addition to the mildly
relativistic particles which are responsible for the formation of the soft X-ray spectrum.
The Compton cross-section for electron
scattering in the presence of a magnetic field was first studied in the
non relativistic limit by \cite{clr71}, and the QED
expression was derived long ago by many authors
\citep{her79,dh86,bam86,hd91}.
However, its form is so complicated to be often of little practical
use in numerical calculations. Moreover, because of their inherent complexity, many of the
published expressions are affected by misprints and the comparison between different derivations
is often problematic.
On the other hand, the use of the full expression of the cross section is especially needed
when a detailed model of cyclotron line formation has to be
evaluated, including expectations for the line profile \citep[see e.g.][]{dv78, ah99}.
In the situation we are considering, it is reasonable to assume that nearly all photons scatter
at resonance. Non-resonant scattering contributions
have negligible effects on shaping the overall spectrum, except
possibly in the very neighborhood of a cyclotron line peak.

Motivated by this, we present here explicit, relatively simple expressions for the
magnetic Compton cross-section at resonance that can be then included in Monte Carlo calculations
such as that described in Paper I.
In doing so, we investigate the behaviour of the different terms and assess their relative importance.
The final expressions have been cross-checked by comparing different
published formulations, when specialized at resonance. A direct application of the
results discussed here to spectral calculations will be presented in a forthcoming study
(Nobili, Turolla \& Zane, in preparation). The paper is
organized as follows. In \S\ref{first} we formulate the problem and
summarize the main ingredients needed for a Monte Carlo simulation. In
\S\ref{Master} we compute the relevant cross sections, specified at
the resonance, while the transition rates (which are related to the
natural width of the excited resonant levels) are given in \S\ref{TranRate}.
The creation of photon via spawning effects is discussed in \S\ref{Spawning},
while \S\ref{abs} contains a brief comparison between scattering and
absorption. In \S\ref{mean} we compute the optical depth, which yields the probability of scattering.
Conclusions follow in \S\ref{conc}.

\section{Formulation of the problem}\label{first}

In this section we briefly outline our approach, discuss our working assumptions and introduce
the mathematical notation. In dealing with relativistic scattering it is convenient to express
all relevant physical quantities in a dimensionless form.
To this end we introduce the dimensionless photon energy and magnetic field strength
\begin{eqnarray}\label{adim}
 \om  & =&  \hbar \omega/m_e c^2\, , \\
  B   & =&   {\cal B}/{\cal B}_{cr} \== \om_B
\end{eqnarray}
where $\omega$ is the photon angular frequency, ${\cal B}$ the
magnetic field and ${\cal B}_{cr} = m_e^2 c^3/e\hbar = 4.414 \times 10^{13}$~G the critical
QED field, at which the fundamental cyclotron energy $\hbar \omega_B$ equals the electron rest mass energy.
Similarly, we denote with $E$ the particle energy expressed in units of $m_e c^2$,
and  with $p$ the component of the particle momentum parallel to the magnetic field, expressed in units
of $m_e c$.

\begin{figure*}
\includegraphics[width=10cm]{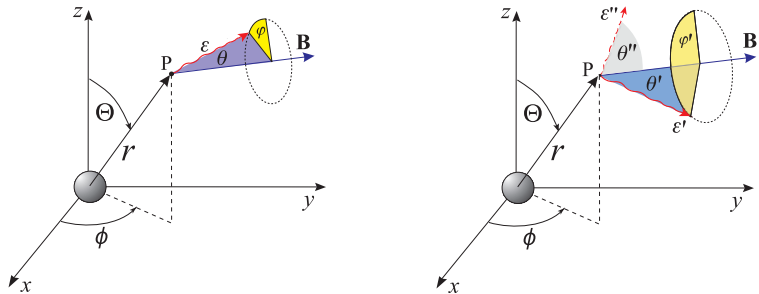}
\caption{Geometry for electron-photon scattering. The left (right) panel refers to
the situation before (after) scattering; all relevant angles are indicated. The direction of the
spawned photon with energy $\om''$ is also shown (note that this is only allowed for intermediate
states with quantum number $n \ge 2$). The figure has only illustrative purposes and here all quantities
are referred to an observer at rest with the star.}\label{Figone}
\end{figure*}

The process of electron-photon scattering can be schematized as follows:
\begin{enumerate}
{\item an electron in the lowest Landau level at a certain position $P$ in
the magnetosphere; the initial spin orientation is necessarily {\it down};}
\item{an incoming photon with energy $\om$ which propagates in a direction forming
angles $\theta$ and $\vphi$ with respect to the magnetic field direction in $P$ (see Figure~\ref{Figone}).
The incoming photon may have either linear\footnote{In general photons propagating in the ``vacuum plus
cold plasma'' are elliptically polarized. However, in our applications, particle density is low enough
($\la 10^{16}\, {\rm cm}^{-3}$) to make the plasma contribution to the dielectric tensor negligible.
The two normal modes are then linearly polarized \cite[e.g.][]{hl06}.} polarization $s=1$,
also indicated by $\parallel$, i.e. parallel to the magnetic field  ({\it ordinary photon\/}) or
$s=2$, also denoted by $\perp$, i.e. orthogonal to $\vec {\cal B}$ ({\it extraordinary photon\/}).}
\item{an excited electron which occupies an intermediate Landau level with quantum number
$n>0$ and energy $E_n$.}
\item{a scattered photon with energy $\om'$, direction $\theta',\ \vphi'$ and polarization $s'=1,2$.}
\item{a recoiled electron in the Landau level $\ell<n$,
with energy $E_\ell = 1+\om-\om'\,$ and parallel momentum $p'$. If $\ell>0 $ the spin of the recoiled electron
can be either {\it up} ({\it spin-flip\/} transition) or {\it down} ({\it no
spin-flip} transition). In the following sections these two possibilities will be denoted, following
\cite{hd91}, with $f=1$ and $f=0$ respectively.}
\item{If $\ell>0$ electron de-excitation to the ground level
is accompanied by the emission of one or more photons (Landau-Raman scattering).}
\end{enumerate}

In order to derive the relevant expressions at
resonance, we start from the fully relativistic magnetic Compton scattering cross section as
derived by \cite{dh86} and \cite{hd91}.
We do not refer to the ultra-relativistic limit discussed in \cite{gon00}, because our final
goal is to derive expressions that can be used in simulations where
different electron populations are present, from non-relativistic to
mildly and ultra-relativistic.
The cross section  derived by \cite{dh86} and \cite{hd91} includes excitation of the electron to
arbitrary Landau levels. However, since we are interested in computing spectra up to
a few hundred keVs, scattering occurs (for all resonances) where $B\la 1$. For such fields the
probability of exciting the second Landau levels is already modest, although non-negligible (see
\S\ref{Master} and Figure~\ref{Figdue}), and becomes even smaller
(typically by a factor $\ga 10$) for $n>2$. For this reason we will assume
that the scattering process occurs only via intermediate states $n\leq 2$ and, consequently, the final level
of the electron (i.e. after the scattering) can be either $\ell=0$ or $\ell=1$. In the latter case  the
electron remains in an excited state and the rapid de-excitation
to the ground level is accompanied by the emission of a new photon.
The relevance of this process (photon spawning) is discussed in Section \ref{Spawning}.

In the following sections we will make use of four different reference
frames: a fixed frame (LAB), centered on the neutron star; a frame in
which the electron has vanishing parallel momentum before the scattering
(the electron rest frame, \ERF); a frame in which the recoiled charge has
vanishing parallel momentum (\ERFS); a frame in which the electron is at
rest in its virtual excited state $n$ (\ERFSS). Unprimed (primed) variables
refer to quantities before (after) the scattering.

\section{Resonant cross sections for photon-electron
scattering}\label{Master}

In strong magnetic fields the transverse momentum of a
charge is not conserved upon scattering since the field can absorb or add momentum
to the particle. The charge momentum parallel to the
magnetic field, $p$, instead,
is related to the parallel components of the photon momentum before and after the collision  by
the usual conservation law, which, in the \ERF\ reads
\begin{eqnarray}\label{emomentum}
 p' &=& \om\ct-\om'\ct'.
\end{eqnarray}
After the collision the particle is left in a state with principal quantum number $\ell\ge 0$ and
energy $E'_\ell \==\sqrt{1+{p'}^2+2\ell B}$. Energy and parallel momentum
as measured in the \ERF\ and in the frame where the recoiled electron
is at rest, \ERFS, are then related by the Lorentz transformations
\begin{eqnarray}\label{EpLorentz}
\sqrt{1+2\ell B} &=& E^*_\ell \== \gamma'\, (E'_\ell - \beta'\, p' )\qquad\quad {\rm and} \qquad\quad
0 \== p^* \== \gamma'\, (p'-\beta'\, E'_\ell),
\end{eqnarray}
from which the velocity and energy of the excited particle in the \ERF\ follows
\begin{eqnarray}\label{velocity}
\beta' &=& \su{p'}{\sqrt{1+{p'}^2+2\ell B}}\,\, ,\qquad
\gamma' \== \su{\sqrt{1+{p'}^2+2\ell B}}{\sqrt{1+2\ell B}}\,\, .
\end{eqnarray}
Finally, parallel momentum and energy conservation yields the final photon energy
\begin{eqnarray}\label{omegap}
\om' &=& \su{\om^2\stq +2\om - 2\ell B }{1+\om (1-\ct\ct')
+ \sqrt{\,1+2\om\ct' (\ct'-\ct) + \om^2 (\ct-\ct')^2 + 2\ell B\stq'}}\, .
\end{eqnarray}

A strong magnetic field greatly affects the scattering cross section, but its
effect is substantially different in the case the photon electric vector
is parallel or perpendicular to $\vec{\cal B}$.
For frequencies below $\omega_B$ the cross section of the extraordinary mode ($\perp$-mode)
is greatly suppressed with respect to the Thomson value, $\sigma_T$
\citep{clr71}. Both polarization modes manifest, however, a singular behavior at nearly
regular frequencies $\omega_n$, which correspond to excitations of the
quantum Landau levels.
In the frame where the electron is initially at rest and using the dimensionless variables defined
in \S\ref{first} the relativistic second order cross section for the transition from the ground
to an arbitrary state $\ell$ is
\begin{eqnarray}\label{crosssection}
\su{d\sigma_{s\to s'}}{d\Omega'} &=& \su{3\sigma_T}{16\pi}\,\su{\om'}{\om}\,\su{(2+\om-\om')\,
\exp{[-(\om^2\stq+{\om'}^2\stq')/2B}]}{[1+\om-\om'- (\om\ct-\om'\ct')\ct']}\,
\left\vert\,\sum_{n=0}^{\infty}\,\left[  F_{n,-}^{(1)} + F_{n,+}^{(1)}+
\left( F_{n,-}^{(2)} + F_{n,+}^{(2)} \right) \exp{(2i\Phi)}\right]\,\right\vert^2
\end{eqnarray}
(see equation [11] of \citealt{hd91} where some misprints have been corrected).
Here $d\Omega'=\sin\theta'd\theta' d\varphi'$, the phase $\Phi$ depends on the difference $\vphi-\vphi'$, and
the complete expressions for the functions $F_{n,\pm}^{(k)}$ can be found in the Appendix of
\cite{hd91}\footnote{Note that, due to a typo, in equations A1-A5 
of \cite{hd91}, all
terms $N\eps_{.}\eps'_{.}\Lambda_{..}\Lambda_{..}$
must be replaced by $N^2\eps_{.}\eps'_{.}\Lambda_{..}\Lambda_{..}$}.
The major complication in equation (\ref{crosssection}) derives from the presence of
an infinite sum over all intermediate (virtual) states with principal quantum number $n$.
In addition, the $F_{n,\pm}^{(k)}$ are
complicated complex functions of $B,\,\om$, $\om'$, $\theta\,,\theta'$, $\vphi\,,\vphi'$.
They depend also on  the initial and final photon polarization mode, $s,\,s'$, and
on the spin orientation of the electron in the intermediate state, i.e. {\it spin-up} (labelled by the index
$+$) or {\it spin-down} (index $-$). Finally, if $\ell>0$, the form of these functions is different if
the orientation of the electron spin in the final state is {\it up} ($f=1$) or {\it down}
($f=0$). If the final state is the ground state only spin down is allowed.

All the $F_{n,\pm}^{(1)}$ exhibit a divergent behaviour at
the resonant energies
\begin{eqnarray}\label{omres}
\om_n &=& \su{2nB}{1+\sqrt{1+2nB\sin^2\theta}}\, .
\end{eqnarray}
while the $F_{n,\pm}^{(2)}$, which are obtained
from $F_{n,\pm}^{(1)}$ by plane crossing symmetry replacement of variables \cite[see again][]{hd91}, remain
instead finite\footnote{
Actually also the $F_{n,\pm}^{(2)}$ 
diverge, but this occur only at photon energies
$>2/\sin\theta$ 
which is the threshold for pair production and where 
expression (\ref{crosssection}) is not
valid any more (e.g. \citealt{her79})}. Since, as mentioned
earlier on, we are interested only in resonant scattering, this leads
to a major simplification. Neglecting all non-resonant 
contributions in (\ref{crosssection})
when $\om\simeq \om_n$, which amounts to disregard both the $F_{n,\pm}^{(2)}$
and the non-resonant terms in the $F_{n,\pm}^{(1)}$, makes it possible to express the cross section
as a infinite sum of separated terms
\begin{eqnarray}
\su{d\sigma_{s\to s'}}{d\Omega'}  &=& \sum_{n=1}^\infty\su{d\sigma^{(n)}_{s\to s'}}{d\Omega'}
\label{pcross}
\end{eqnarray}
where each resonant contribution is given by
\begin{eqnarray}
\su{d\sigma^{(n)}_{s\to s'}}{d\Omega'} &=& \su{3\sigma_T}{16\pi}\,\su{\om'}{\om}\,
\su{(2+\om-\om')\,\exp{[-(\om^2\stq+{\om'}^2\stq')/2B]}}{[1+\om-\om'- (\om\ct-\om'\ct')\ct']}
\, \left\vert\, F_{n,+}^{(1)}\++ F_{n,-}^{(1)} \,\right\vert^2\label{pcrossn}.
\end{eqnarray}

The quantities  $F_{n,\pm}^{(1)}$ are independent on $\vphi$ and have the general form
\begin{eqnarray}\label{FunF}
  F_{n,\pm}^{(1)} &=& \left(\su{1+E_n}{2E_n}\right)
\, \su{\Tn{\pm} (\om,\om', \theta, \theta',B;\, s,s',f)}{1+\om-E_n}
\end{eqnarray}
where
\begin{eqnarray}\label{Enelectron}
  E_n &=& \sqrt{1+\om^2\ctq+2nB}
\end{eqnarray}
is the energy of the electron in the intermediate state. As discussed in
\S\ref{first}, we restrict our treatment to situations in which scattering
occur only via excitation of intermediate states $n \leq 2$, i.e. we only
consider the first two resonant terms in equation~(\ref{pcrossn}); the
corresponding expression for $\Tn{\pm}$ are given explicitly in the Appendix.

Actually, the divergence of the $F_{n,\pm}^{(1)}$, which
occurs when the denominator $1+\om -E_n$
vanishes, reflects an unphysical behaviour of the cross section and is
merely a consequence of the fact that the expressions presented so
far have been computed without accounting for the finite
lifetime of the electron in the excited Landau levels.
In a realistic situation, according to the Heisenberg principle,
each excited level has an energy indeterminacy which is
proportional to the inverse of the lifetime of the electron in that state, i.e. to the
decay rate.
This is standardly accounted for by introducing
the natural line widths, which amounts to perform the substitution
\begin{eqnarray}
  E_n \ &\to & \ E_n - {i}\Gn\pm/2,
\end{eqnarray}
in the denominator of equation (\ref{FunF}); here the $\Gn\pm$ are related
to the relativistic decay rate out of the $n$-th intermediate state
and their explicit expressions will be presented in \S \ref{TranRate}. With this
substitution, the functions (\ref{FunF}) become
\begin{eqnarray}\label{FnLorentz}
F_{n,\pm}^{(1)} &=& \left(\su{1+E_n}{2E_n}\right)\, \su{\Tn{\pm}}{1+\om-E_n + i\Gn{\pm}/2}.
\end{eqnarray}
A more useful representation is obtained
by replacing the Lorentzian that naturally arises from (\ref{FnLorentz}),
with a $\delta$-function. By taking the limit
\begin{eqnarray}\label{Lorentian}
 \lim_{\Gn\pm\to 0} \su{ \Gn\pm /2\pi }{(\om-E_n+1)^2+ (\Gn\pm/2)^2} &=&  \delta (\om-E_n+1)
\end{eqnarray}
we obtain, after a lengthy but straightforward calculation,
\begin{eqnarray}\label{ResTerm}
  \left\vert\, F_{n,+}^{(1)} + F_{n,-}^{(1)}\,\right\vert^2  & = &
\su{\pi}{2} \, \left( \su{1+E_n}{E_n} \right)^2 \, \left[\su{({\Tn+})^2}{\Gn+} + \su{({\Tn-})^2 }{\Gn-} +
\su{ 4\, \Tn+ \, \Tn- }{\Gn+ +\Gn-}\right] \su{E_n(\om_n)\, \delta(\om-\om_n)}{\sqrt{1+2nB\sin^2\theta}}\, .
\end{eqnarray}
We note that the factor 
$E_n(\om_n)/\sqrt{1+2nB\sin^2\theta}$ in eq. (\ref{ResTerm}) arises because 
of the change in the argument of the $\delta$-function, 
from $\om -E_n(\om)+1$ to $\om-\om_n$. Ultimately
it reflects the fact that the ``effective'' width of the scattering line profile near resonance is
$\Gamma_nE_n(\om_n)/(2\sqrt{1+2nB\stq})$ \cite[see][]{hd91}.

Substituting back equation~(\ref{ResTerm}) into equation (\ref{pcrossn})
and upon a trivial integration on $\vphi'$, brings the $n$-th resonant term
(\ref{pcrossn}) in the form
\begin{eqnarray} \label{npcross}
\su{d\sigma^{(n)}_{s\to s'}}{d{\ct\,}'} & = & {\cal D}^{(n)}_{s\to s'}(B,\theta,\theta')\, \delta (\om-\om_n),
\end{eqnarray}
where
\begin{equation}\label{npcrossa}
{\cal D}^{(n)}_{s\to s'}= \su{3\pi\sigma_T}{16}
\su{(1+E_n)^2}{E_n \sqrt{1+2nB\sin^2\theta} }\,\su{\om'\,(2+\om-\om')
\exp{[-(\om^2\stq+{\om'}^2\stq')/2B]}}{\om[1+\om-\om'-(\om\ct-\om'\ct')\ct']}
\left[\su{({\Tn+})^2}{\Gn+}+\su{({\Tn-})^2 }{\Gn-}+\su{ 4\,\Tn+\,\Tn-}{\Gn+ +\Gn-}\right]\, ,
\end{equation}
and $\om'$ is given by equation (\ref{omegap}). In the previous expression one can
safely put $\om=\om_n$ since ${\cal D}^{(n)}_{s\to s'}$ in multipled by $\delta (\om-\om_n)$.

\begin{figure*}
\includegraphics[width=17cm, height=6.6cm]{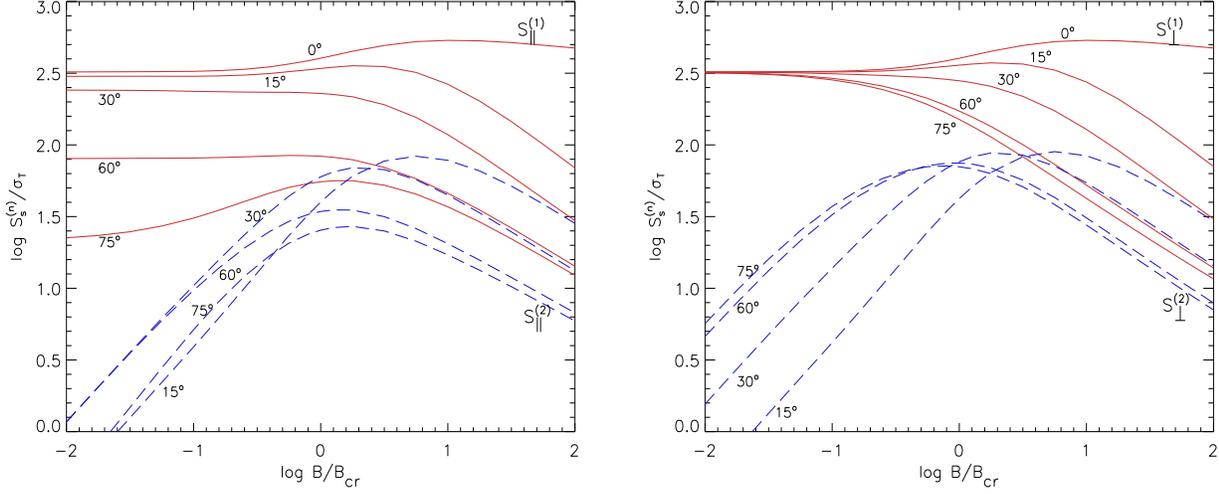}
\caption{The resonance factors for the first (upper curves, solid red lines) and second (lower curves, dashed blue lines)
intermediate Landau levels as a function of $\log B$; the $S^{(n)}_s$ are in unit of the Thomson cross section.
Different curves are labelled by the value of the angle between the incident photon direction and the magnetic
field. {\it Left:} ordinary photons.
{\it Right:} extraordinary photons (note the weaker angular dependence in
the latter case). Clearly, for a photon propagating parallel to the magnetic field only the
first resonance exists and  $S^{(2)}_s$ vanishes.}
\label{Figdue}
\end{figure*}

The $n$-th resonant term in the angle-integrated cross section is
\begin{eqnarray}
\label{Icross}
 \sigma^{(n)}_{s\to s'} &=& {S}^{(n)}_{s\to s'} (B,\theta; \ell, f)\delta(\om-\om_n) \, ,
\end{eqnarray}
where the integral
\begin{eqnarray}\label{IcrossT}
{S}^{(n)}_{s\to s'} (B,\theta;, \ell, f) & = & \int_{-1}^{1} {\cal D}^{(n)}_{s\to s'} d({\ct\,}')
\end{eqnarray}
can be evaluated numerically.
The $n$-th resonant term of the {\it total \/} cross section is then
obtained by summing equation~(\ref{Icross}) over the polarization states of
the scattered photon, $s'$, and over the energy
levels $\ell$, and the spin states, $f$, of the recoiled electron.

When $n=1$, the terms $\, S^{(1)}_{s\to s'}$ and ${\cal D}^{(1)}_{s\to s'}\, $
are not associated with a choice of different quantum states $\ell$ and
$f$, because in this case it necessarily follows that $\ell=0 $ and $f=0$.
Therefore, for a photon with given initial polarization $s$, the first resonant term
of the {\it total\/} cross section is simply given by the sum of two contributions, i.e.
\begin{eqnarray}\label{cross1}
 \sigma^{(1)}_{s} (\om,B,\theta) & = & \left[S^{(1)}_{s\to 1} (B,\theta;\, \ell=0, f=0) \++
  S^{(1)}_{s\to 2} (B,\theta;\, \ell=0,f=0) \right]\delta (\om-\om_1) \==
S^{(1)}_s(B,\theta) \, \delta (\om-\om_1).
\end{eqnarray}
Instead, the second resonant term of the {\it total\/} cross section,
corresponding to the intermediate state $n=2$, must account for
both final states $\ell=0,\, 1$ and, if $\ell=1$, for two different
orientations of the electron spin. Its expression is then 
\begin{eqnarray}
\nonumber
\sigma^{(2)}_{s} (\om, B,\theta) & = &
\left\{\left[S^{(2)}_{s\to 1} (B,\theta; \ell=0,f=0) \++ S^{(2)}_{s\to 2}(B,\theta; \ell=0,f=0)
\right] \++ \right. \\
&& \left.\left[S^{(2)}_{s\to 1} (B,\theta; \ell=1,f=0)\++ S^{(2)}_{s\to 2} (B,\theta; \ell=1,f=0) \right] \++
\right.\\
\nonumber
&&\left. \left[S^{(2)}_{s\to 1}(B,\theta; \ell=1,f=1)+S^{(2)}_{s\to 2}(B,\theta; \ell=1,f=1)\right]
\right\}\delta (\om-\om_2)\== S^{(2)}_s(B,\theta)\, \delta (\om-\om_2)\, .
\label{cross2} \end{eqnarray}
Finally, the total cross section is obtained by adding the different resonant terms
\begin{eqnarray}\label{TotalCross} \sigma_s (\om, B,\theta) &=&
\sigma^{(1)}_{s} (\om, B,\theta) + \sigma^{(2)}_{s} (\om, B,\theta) =
\sum_{n=1}^2 \, S_s^{(n)} (B, \theta)\,\delta (\om-\om_n).
\end{eqnarray}
The eight functions $\, S^{(n)}_{s\to s'}\, $ which appear in
equations~(\ref{cross1})-(\ref{cross2}) are the fundamental ingredients for
any computational investigation of photon diffusion in strong magnetized plasmas.
In particular, when performing Monte Carlo simulations, proper
combinations of these expressions can be used to evaluate the probability of transition
between different states.

Figure \ref{Figdue} illustrates the dependence of the functions
$S^{(1)}_{s}$ and $S^{(2)}_{s}$ on the magnetic field strength for both
polarization states and different incident photon direction. The
non-relativistic limit 
tangent to the low $B$-field limit of each curve and 
takes the form
\begin{eqnarray}\nonumber 
S_\perp^{(1)} &\sim& \su{3\pi}{8\alpha}(2+B\ctq)\,, \hskip1.cm S_\parallel^{(1)} \sim \su{3\pi}{8\alpha}(2\ctq+B)\cr
\nonumber
&&\cr
\nonumber
S_\perp^{(2)} &\sim& \su{3\pi}{2\alpha}B\stq\,, \hskip1.775cm S_\parallel^{(2)} \sim \su{3\pi}{2\alpha}B\stq(B+\ctq)\,,
\end{eqnarray}
where $\alpha=1/137$ is the fine structure constant.
In the ultra-relativistic limit no simple analytical expansion can be derived. Approximate expressions, which hold for
$\theta\neq 0$,  are
$S^{(1)}_\parallel\simeq S^{(1)}_\perp\propto 1/(1+2\sqrt{B\stq }\,)$ and $S^{(2)}_\parallel\simeq S^{(2)}_\perp\propto 
1/(1+\sqrt{B\stq }\,)$. As it can be seen,
relativistic corrections are already $\sim 10\%$ at $ {\cal B} \sim 0.1\,
{\cal B}_{cr}$ for photons with large values of the
incident angle $\theta$ (see Figure \ref{Figdue}).
Furthermore, we may note that the number of excitations of the second
Landau level, totally negligible in the weak $B$ limit, becomes sizable in
moderately relativistic regimes.

This result is more evident in Figure \ref{RapportiLFinal} (left panel), where we show the
ratio between the second resonant term and the total cross section in the case of
unpolarized incident photons, obtained averaging over ordinary and
extraordinary modes ($s=1,\, 2)$. As it can be seen, in strong magnetic fields a non-negligible
fraction of collisions can excite electrons to the higher Landau level ($n=2$).
It is also worth noticing that, when $n=2$, de-excitation of the electron to $\ell=1$ state
is generally more likely than the direct transition to the ground level, as it can be seen from
the right panel in Figure~\ref{RapportiLFinal}. Only for magnetic field strengths largely exceeding
the critical quantum limit, the two transitions are have comparable probabilities.
This has important consequences on the transfer problem, since it implies that collisions
involving the intermediate state $n=2$ lead more frequently to the creation of extra photons
as the electron returns to the ground level.
In other words, the resonant magnetic scattering may not conserve the total photon number.

\def\figd{7.5cm}
\begin{figure*}
\includegraphics[width=\figd]{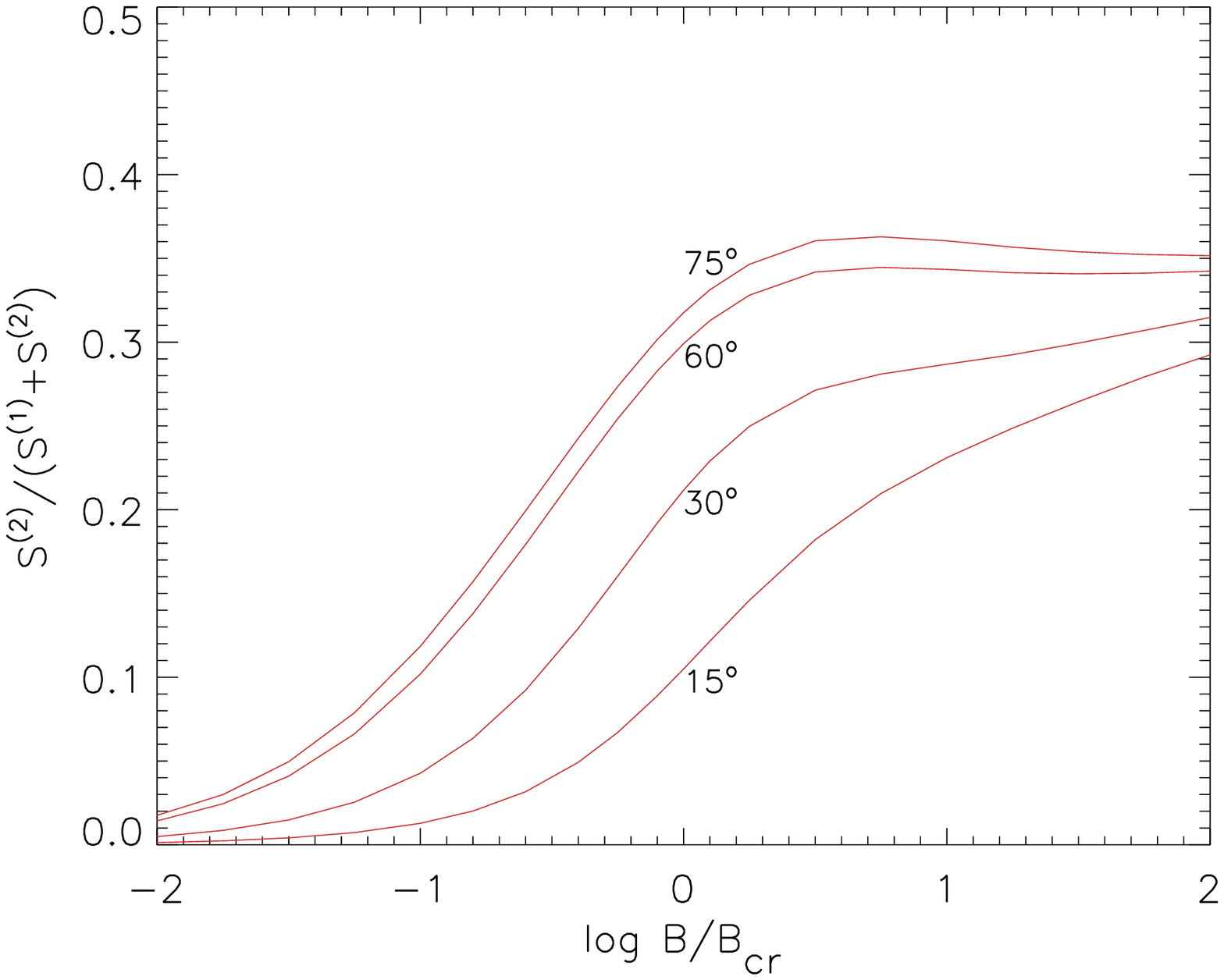}
\hskip5mm
\includegraphics[width=\figd]{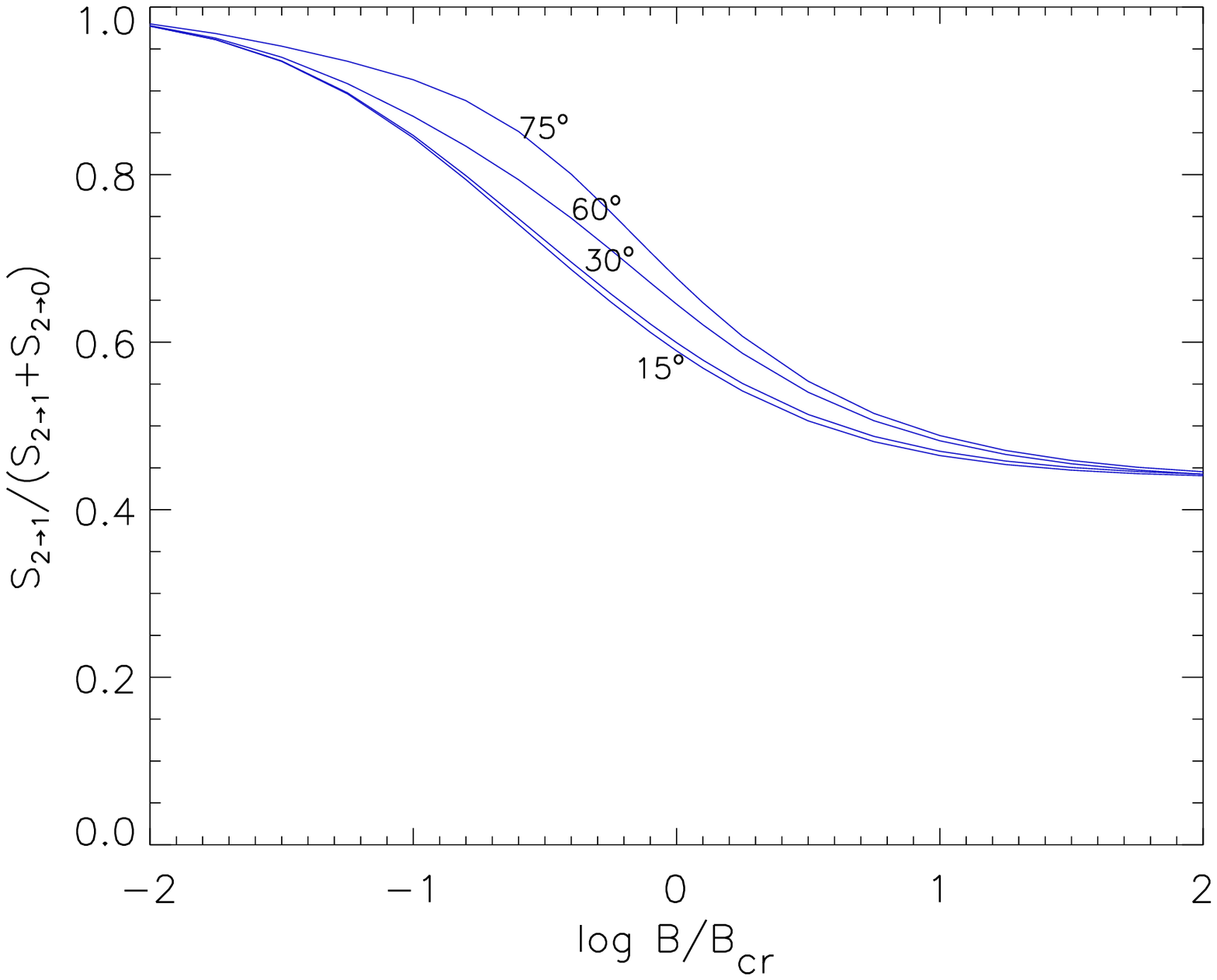} \
\caption{{\it Left:\/} The ratio  $S^{(2)}/( S^{(1)}+S^{(2)})$ versus the
magnetic field strength for three different scattering angles. The functions $S^{(n)}$
are obtained by summing over the two initial photon polarization states, i.e.
$S^{(n)}=S^{(n)}_1+S^{(n)}_2$.
{\it Right: \/} The $2\to 1$ transition probability versus magnetic field
strength.
}\label{RapportiLFinal}
\end{figure*}

\section{Transition Rates}\label{TranRate}

For the sake of completeness, in this section we present the explicit
expressions for the transition rates and for the natural line widths
$\Gn\pm$. The differential transition rates (number of decays per second
and per sterad) for electrons with initial spin {\it up} ($+$) or {\it
down} ($-$) were derived in a general form by \cite{dv78} and \cite{hrw82} 
\cite[see also][]{la86,bgh05}.
Here we will use the derivation presented by \cite{hrw82}, by focussing
only on those cases which are of immediate relevance to the present investigation.
In this respect, we note that, while performing a Monte Carlo simulation,
the transitions rates have a twofold role. First, as discussed in \S~\ref{Master}, they
fix the natural width of the Landau levels, and therefore enter directly into the numerical
computation of the resonant scattering cross section. Second, as we will see in \S~\ref{Spawning},
they determine the properties of the photons emitted by successive
de-excitations when, in Landau-Raman scattering the electron is left in an 
excited state $\ell >0$.
In practice, in the particular case we are discussing (i.e. only levels
up to $n=2$ can be excited), it is the differential transition rate $d R_{\pm}^{\, 1\to 0}
/d\Omega$ (see equation~\ref{Transn} below) that fixes the properties of the photon which is emitted
through the decay of an electron from $\ell=1$ to the ground level.

The cross sections discussed in the previous sections have been computed
in the frame in which the electron is at rest before scattering (ERF, see
\S\ref{Master}), and the parallel component of the momentum in the
intermediate state $n$ is $p=\om_n\ct$. However, in this section we
find more convenient to proceed by first introducing the transition rates
and line widths $\,\widetilde{\Gamma}_{\pm}^{n}\,$ as computed in the reference frame \ERFSS\,
in which the electron is momentarily at rest in the virtual state $n$, and
then to transform them back in the \ERF. In fact, in the \ERFSS\ the equations governing the
de-excitation and the consequent photon emission take their simpler form. Furthermore, in this
section we will indicate with $\, E_n\, $ and $\, E_\ell\,$ the energy of the electron in the state
$n$, $\ell$, and with $\ome$ and $\theta_e$
the energy and direction (wrt the magnetic field) of the emitted photon,
respectively. All these quantities are computed in the \ERFSS.

In the \ERFSS, the energy of an electron which is excited to the level $n$ is therefore
\begin{eqnarray} \label{Eemission}
 E_n &=& \sqrt{1+2nB}\, .
\end{eqnarray}
By applying again energy and parallel momentum conservation we obtain the energy of the electron
de-excited to the level $\ell$ and that of the emitted photon
\begin{eqnarray}
\label{Emisfe}
 E_\ell &=& E_n \-- \ome \== \sqrt{1+2\,\ell\,B + \ome^2\ctqe}\\
\label{Emisfp}
 \ome   &=& \su{2(n-\ell)B }{E_n+\sqrt{1+2B (n\ctqe+\ell\stqe)}}
\end{eqnarray}
respectively.

The differential rate for transitions between a generic level $n>0$ and
the ground state $\ell=0$, as given by \cite{hrw82}, takes the form
\begin{eqnarray}\label{Transn}
\nonumber
\su{d R_{\pm}^{\, n\to 0} }{d\Omega} & = & \su{\alpha mc^2}{4\pi\hbar}\,\su{\ome\, \exp{(-Z)}\,
Z^{n-1}}{(n-1)!\, E_n\,(E_n-\ome\stqe)}\left\{\su{Z}{n}\,\left[\su{}{}
2nB-\ome\,(E_n\,\mp\,1)\right]\langle\stqe,0\rangle \++ \right. \\
&& \left. \++ \left[\su{}{} 2nB-\ome\,(E_n\,\pm\,1)\right]\,\langle\ctqe,1\rangle
\++ 2\om^2\,\ctqe\langle\stqe,0\rangle\right\},
\end{eqnarray}
where $Z=\ome^2 \stqe/2B$.
The angular factors in ``$\langle\,\rangle$'' refer to two possible linear polarization states
$s=1,2$. The rate for transition from  $n=2$ to $\ell=1$ is
\begin{eqnarray}\label{Trans1}
\nonumber \su{d R_{\pm}^{2\to 1}}{d\Omega} &=& \su{\alpha mc^2}{4\pi\hbar}
\,\su{\ome \, \exp{(-Z)} }{\, E_n\, (E_n-\ome\stqe\, ) \, } \left\{Z\, \left[[\su{}{} 4 B-\ome\,
\,(E_n\, \mp \, 1)]{(Z-2)^2}/{2}\++\right.\right. \\
\nonumber && \left. +\ \su{}{}4B-\ome\,(E_n\,\pm\,1)+4B\,(2Z-3)\right]\,\langle{\stqe},0\rangle
\++\left[\su{}{}[\,4B-\ome\,(E_n\,\pm\,1)]\,(Z-1)^2\++\right. \\
&&\left. +\ [\su{}{}4B-\ome\,(E_n\,\mp\,1)]\, Z^2/2\right]\,\langle\ctqe,{1}\rangle
\++\left. 2\ome^2\,\ctqe (Z^2-2Z+2)\,\langle{\stqe,\,0}\rangle\su{}{}\right\}.
\end{eqnarray}
Since we are considering only the first two resonances, the two previous expressions are all we
need in order to evaluate the natural widths and photon emission following Landau-Raman scattering
(see \S\ref{Spawning}).

Total rates can be obtained by integrating over the solid angle and summing over
the two polarization states $s$ of the emitted photon, and in general take the form
\begin{eqnarray}\label{TotRate}
R_{\pm}^{n\to\ell} &=& 2\pi\int_{-1}^1 \,\sum_{s=1}^2\,\su{d R^{n\to\ell}_{\pm}}{d(\cte)}\, d(\cte)\, .
\end{eqnarray}
The reciprocal of the transition rate gives the mean lifetime of the
electron in the excited Landau level. Then, due the Heisemberg principle,
each term $R_{\pm}^{n\to\ell}$ is proportional to the energy width for the
particular transition $n\to\ell$.
The total energy width of the $n$-th level is therefore obtained by
summing over all contributions associated with level $n$
\begin{eqnarray}\label{TTRate}
  \widetilde{\Gamma}_{\pm}^{n} &=& \su{\hbar}{m_e c^2}\sum_{k=0}^{n-1}R^{n\to k}_{\pm}\,.
\end{eqnarray}

\begin{figure*}\label{Widths}
\includegraphics[width=7cm]{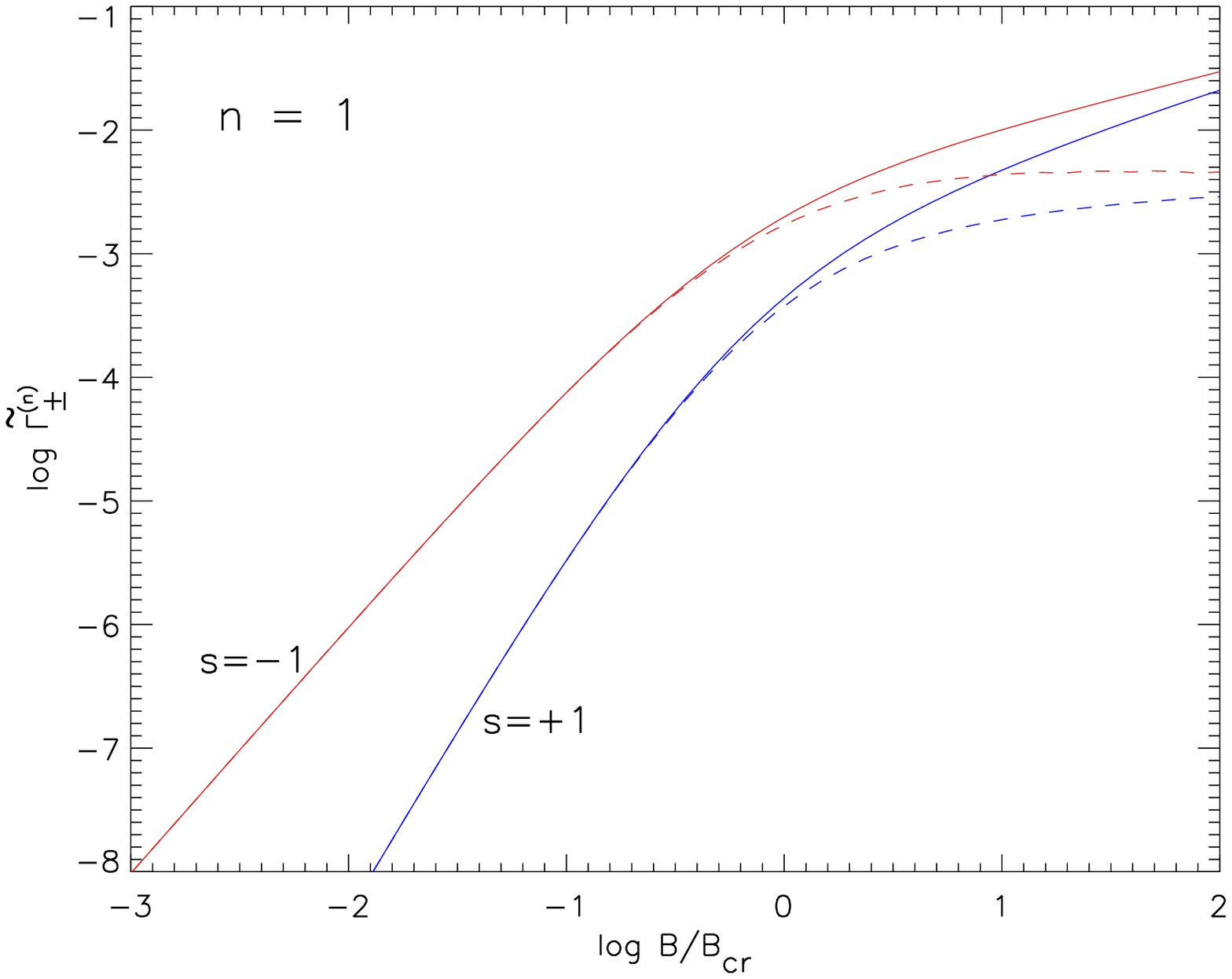}
\includegraphics[width=7cm]{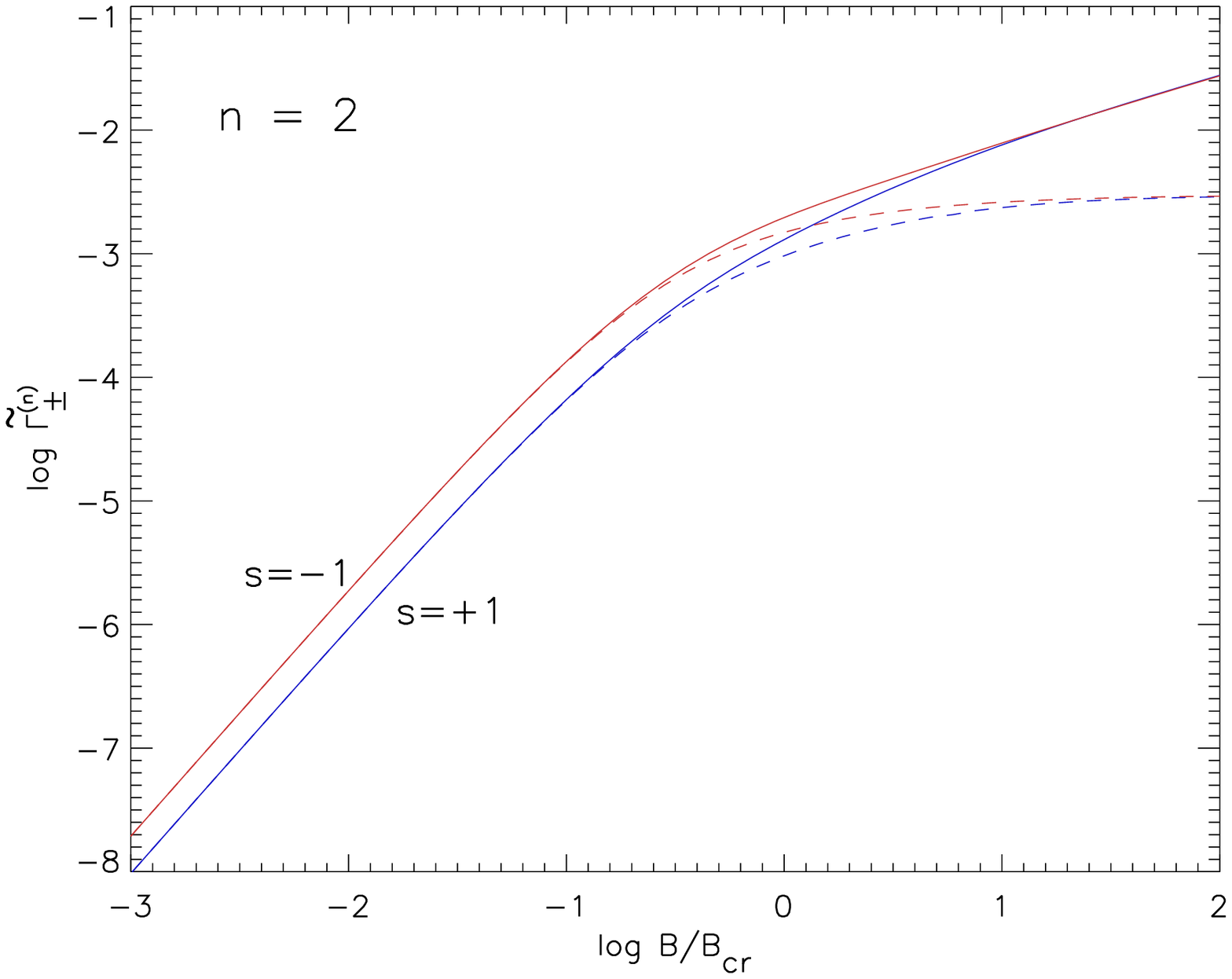}
\caption{Energy widths of the Landau levels (in units of $m_e c^2$) for the two spin orientations
as a function of the magnetic field strength. {\it Left:} $n=1$.  {\it Right:} $n=2$. In each panel
the full lines refer to the widths as computed in the excited electron frame [eq. (\ref{TTRate})] and
the dashed lines to the same quantities but referred to the ERF [eq. (\ref{ATRate})] for $\cos\theta=\pm 1$,
i.e. in the case in which the incoming photon is parallel to $\mathbf B$. }
\label{FigWidth}
\end{figure*}

As previously stated, these expressions are valid in the reference frame in which the
excited electron is at rest, \ERFSS. In order to  obtain the energy widths $\Gn{\pm}$ that
must be used in the expressions given in \S~\ref{Master}, we need to perform a
Lorentz transformation back to the \ERF. This gives
\begin{eqnarray}\label{ATRate}
  \Gn{\pm} &=&\su{\sqrt{1+2nB}}{\sqrt{1+\om_n^2\ctq+2nB}} \,\,\widetilde{\Gamma}_{\pm}^{n}\, ,
\end{eqnarray}
where $\om_n$ and $\theta$ are the incident photon energy (equation [\ref{omres}]) and direction.
Figure \ref{Widths} shows the dependence on the magnetic field strength of the
total energy widths for the first two Landau levels, $\widetilde{\Gamma}_{\pm}^{1,2}$, in the range
$10^{-3} \leq {\cal B}/{\cal B}_{cr} \leq 10^2$. For weak fields they are proportional to $B^2$  
(in particular $\widetilde\Gamma^{1}_- \sim 4\alpha B^2/3$),
except for the spin flip transition for which $\widetilde\Gamma^{1}_+ \sim 2\alpha B^3/3$  
\cite[e.g.][]{hrw82,la86}.
Similarly to what happens for the cross sections, remarkable deviations from the simple, non-relativistic
power-law dependence appears at $B> 0.1$. We note that in the strong field limit all energy widths have a similar
dependence on the field strength, $\sim 0.4 \alpha\sqrt{B}$, a result similar to
that found by \cite{la86} for the total transition rate. 
The  
limiting behaviours of $\widetilde{\Gamma}_{\pm}^{1}$ can be usefully 
checked against those derived by \cite{bgh05} who obtained an analytical 
form (in terms of a series of elementary functions) for the spin-dependent 
transition rates $R^{n\to 0}_{\pm}$. In fact, for $n=1$, it follows from 
eq. (\ref{TTRate}) that 
$\widetilde{\Gamma}_{\pm}^{1}=(\hbar/m_ec^2)R^{1\to 0}_{\pm}$.
As noticed by \cite{bgh05}, their weak field limit coincides with those derived (analitycally) by 
\cite{la86} and discussed above. In the ultra-relativistic limit, on the other hand, one gets
$\widetilde\Gamma^{1}_-\sim\widetilde\Gamma^{1}_+
\sim \alpha(1-1/\mathrm e)(B/2)^{1/2}\sim 0.45\alpha\sqrt{B}$, close to the $B\gg 1$ limit given above.
A direct numerical comparison of our result for $\widetilde{\Gamma}_{\pm}^{1}$ with that of \cite{bgh05}
shows that the fractional difference is below 10\% for $B<100$, with the largest deviations appearing at
large $B$.

\section{Photon spawning}\label{Spawning}

As discussed in the previous sections, some scatterings occur
at the second resonance and excite the electron in the intermediate state $n=2$.
In order to have a significant number of second harmonic excitations
the magnetic field must be sufficiently intense, say
$B\ \gsim \ 0.05$ (${\cal B}
\gsim 2.2 \times 10^{12}$~G; see Figure~\ref{RapportiLFinal} ).
Moreover, the incoming photon must have an energy  $\om=\om_2 \approx 2B$. The two conditions together
require that a significant number of photons with
energy $\hbar\omega \ga50$ keV are present in the magnetosphere.
We do not expect such high-energy photons to be emitted directly by
the stellar surface, which temperature, as inferred by observed X-ray
spectra, is $\sim 0.5$--1 keV, but it must be noted that,
if electrons are relativistic, the photon energy as seen by the moving
particle will be amplified by a factor $\sim\gamma$. Besides,
repeated collisions populate the hard tail of the spectrum and may provide
photons of the required energy.


Once an electron is excited to the second Landau level, it is more likely
left after scattering in the $\ell=1$ rather than in the  $\ell=0$ state
(see \S~\ref{Master} and Figure~\ref{RapportiLFinal}).
This implies that an additional photon will be produced by the prompt radiative de-excitation $1\to 0$.
In general, the transition rate depends also on the rate at which collisions populate the $\ell=1$ level.
The excitation rate due to Coulomb collisions is \citep{bhp79}
\begin{eqnarray}
\nonumber
R_{coll} &\approx&  3\times 10^{10}\left(\su{n_e}{10^{21} {\rm cm}^{-3}}\right)T^{-3/2}\ {\rm s}^{-1}
\quad\approx\quad  1.6\times 10^{6}\left(\su{n_e}{10^{21} {\rm cm}^{-3}}\right)\,B^{-3/2}\ {\rm s}^{-1}\,,
\end{eqnarray}
where the second (approximate) equality follows from the requirement that it has to be $kT\sim \hbar\omega_B$
for the collision to excite the first Landau level. As noticed by \cite{ah99}, in a strongly magnetized,
low-density plasma $R_{coll}$ is negligible when compared to the radiative cyclotron rate

\begin{eqnarray}
R_{rad}  &\approx &  10^{18}\, B^q\ {\rm s}^{-1}
\end{eqnarray}
where the exponent $q$ varies in the range 1/2--2 (see \S\ref{TranRate}).
The population of the excited Landau levels is, therefore, solely regulated in the present case
by magnetic Compton scattering. The large value of $R_{rad}$ justifies the assumption
that electrons remain in the ground level. In fact, the typical time between two scatterings is $\approx
(n_e\sigma c)^{-1}\approx L/(\tau c)\approx 10^{-5}\, {\rm s}\gg 1/R_{rad}$, where the lengthscale $L$ is
a few star radii and the scattering depth $\tau\approx 1$ (see paper I). In a sense, one may view Raman
scattering at the second
resonance, and the ensuing, quick radiative de-excitation to the $\ell=0$ level
(photon spawning) as more akin to double Compton
scattering \cite[e.g.][]{gou84} rather than to absorption followed by
the emission of two photons.

As discussed in \S~\ref{TranRate}, the de-excitation of an electron which is initially at rest in the
$\ell=1$ level with energy  $E = \sqrt{1+2B}$ produces a photon with energy
\begin{eqnarray}\label{emission1}
\ome &=& \su{E-\sqrt{{E}^2-2B\stq_e }}{\stq_e}\==\su{2B}{\sqrt{1+2B} + \sqrt{1+2B\ctq_e }}\, ,
\end{eqnarray}
as computed in its rest frame. The rate at which photons are emitted is given by equation
(\ref{TotRate}) with $n=1,\ \ell=0$, and, when performing a Monte Carlo
simulation, the differential form (\ref{Transn}) can be used to derive, on
a probabilistic ground, the direction of the emitted photon (i.e. the angle
$\theta_e$, emission is isotropic in $\phi_e$). Once again, we stress
that care must be payed to the frame in which these quantities are evaluated.
In fact, the emission angle $\theta_e$ and the photon energy $\ome$ introduced above,
are both referred to the rest frame of the recoiled electron, while in applications they
need to be evaluated in the stellar frame (LAB). This is readily done by means of a Lorentz transformation.
In this respect, we note that the parallel momentum of the recoiled electron is in the LAB
\begin{eqnarray}\label{recoil}
  p' &=& \gamma\beta + \om\,\mu - \om'\,\mu'
\end{eqnarray}
where $\om$ and $\mu$ are the photon energy and the cosine of
the propagation angle with respect to the magnetic field, both measured  in the stellar frame, and,
again, unprimed (primed) variables refer to quantities before (after) the scattering. The corresponding
velocity and the Lorentz factor are
$\,\beta' = p'/\sqrt{1+{p'}^2+2B} \, $ and $\,\gamma' = \sqrt{1+{p'}^2+2B}/\sqrt{1+2B}\, $ respectively
(see \S \ref{Master}), from which we obtain
\begin{eqnarray}\label{recoilangle}
 \cos \theta_e^{(L)} &=& \su{\cte + \beta'}{1+\beta'\cte}
\end{eqnarray}
and
\begin{eqnarray}\label{recoilEn}
  \om_e^{(L)} &=& \gamma'\ome \left(1+\beta'\cte \right)\, ;
\end{eqnarray}
here the index $(L)$  denotes the photon energy and propagation angle as measured in the stellar frame
to distinguish them from the same quantities but referred to  the recoiled electron frame.

\begin{figure*}
\includegraphics[width=8cm, height=6.3cm]{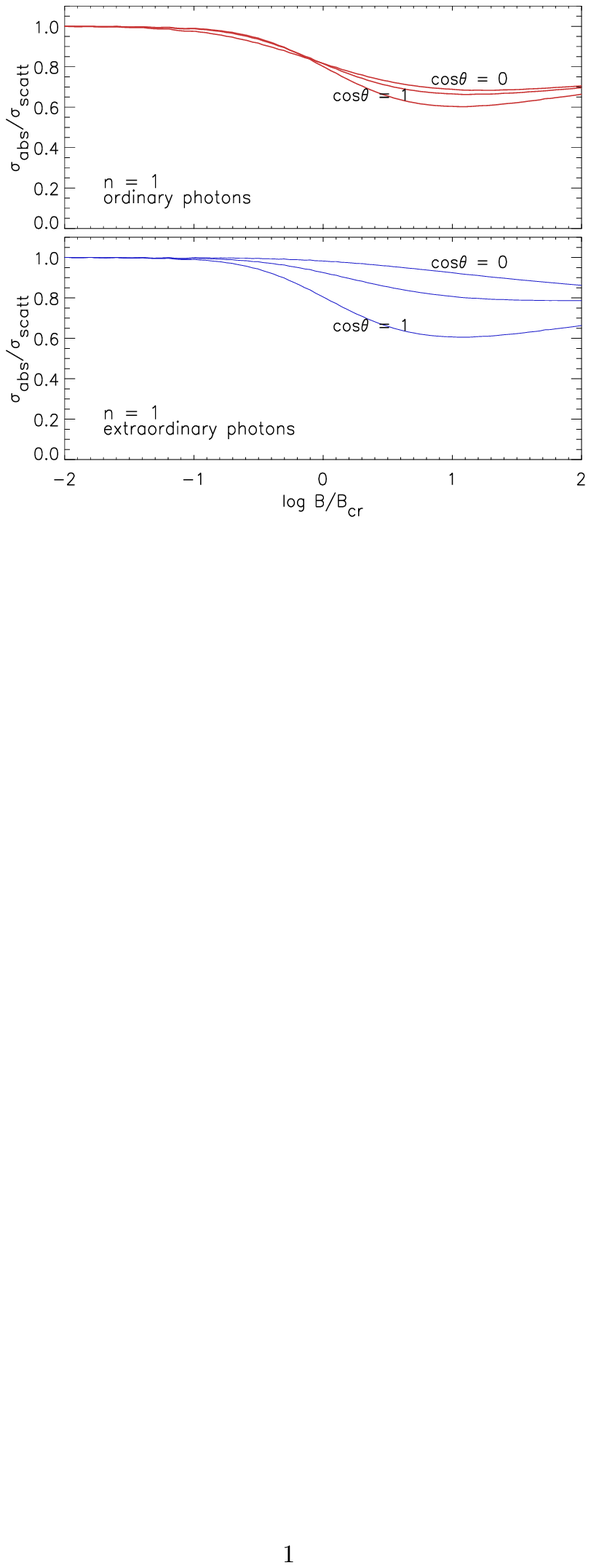}
\includegraphics[width=8cm, height=6.3cm]{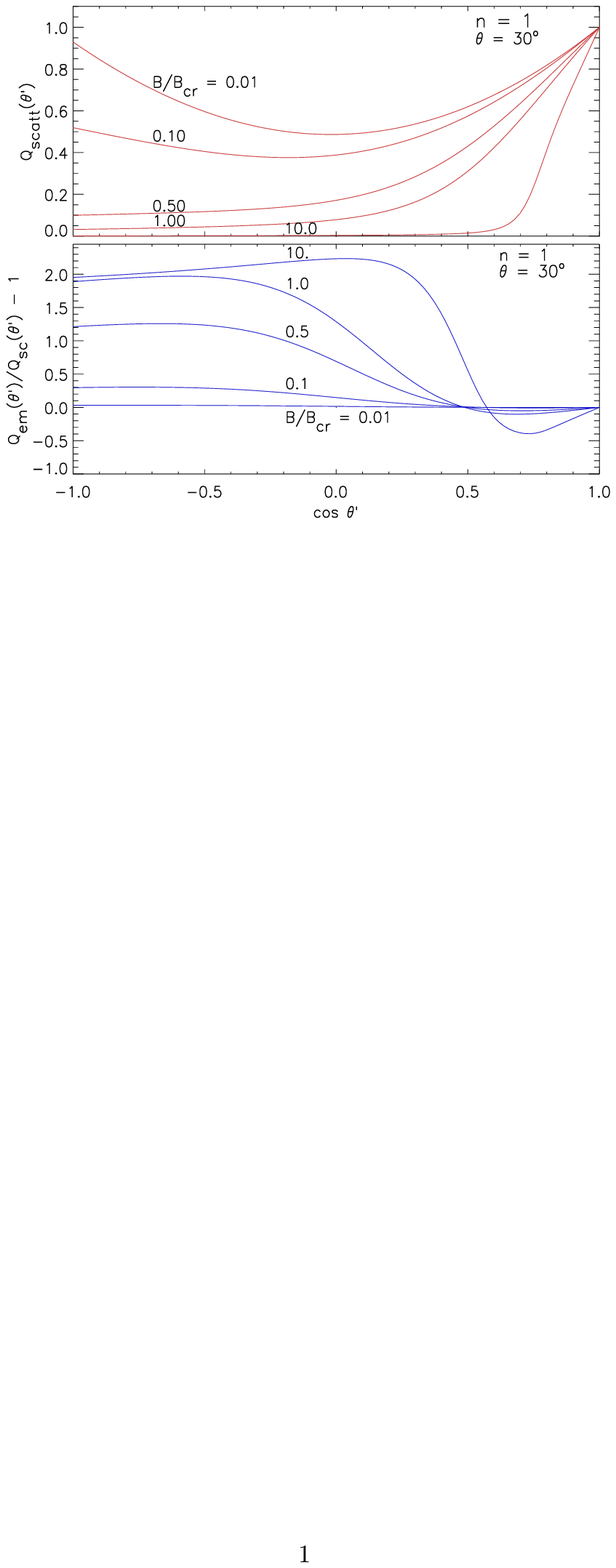}
\caption{{\it Left:} the ratio $\sigma_{abs}/\sigma_{s}$
versus the magnetic field strength for
ordinary (upper panel) and extraordinary (lower panel) photons,
and three different directions of the incident
photon,  $\ct =0,0.5, 1$.
{\it Right:} the angular distribution of scattered photons (upper panel) for a given incident
photon direction $\theta = 30^\circ$. The lower panel shows the fractional difference
between emission and scattering, $(Q_e-Q_s)/Q_s$.
Each curve is labelled by the magnetic field strength.
}\label{Angular}
\end{figure*}

\section{Absorption-emission vs. scattering}
\label{abs}

As noted by \cite{hd91} and \cite{ah99}, in the non-relativistic regime
the on-resonance second order scattering process  $\,\gamma + e \to \gamma'+ e'\, $
is well approximated by the sequence of  two separated first order processes: photon
absorption, $\, \gamma+e\to e^*$, immediately followed by photon emission $\, e^*\to e'+\gamma'$.
Because of its much simpler form, one would like to use the cyclotron absorption cross section
instead of the cumbersome QED scattering cross section in numerical work. It is therefore of interest
to explore the conditions under which  absorption plus emission reproduces relativistic resonant scattering
with reasonable accuracy.

In the \ERF, summed over the final spin states, the angle-averaged absorption cross section can be
written as \citep{dv78}
\begin{eqnarray}\label{absorption}
\sigma_{abs}^{(n)}({\parallel,\perp}) &=& \su{3\pi\sigma_T}{4\alpha} \su{\exp{(-Z)}Z^{n-1}}{(n-1)!
\sqrt{1+2nB\stq}}\left[1\--\su{\om_n}{nB}\, \langle\stq,0\rangle\right] \delta{(\om-\om_n)},
\end{eqnarray}
where the angular factor $\langle\stq,0\rangle$ refers to the two linear polarizations ($\parallel,\ \perp$)
of the incident photon, $Z = \om^2_n \stq/2B$, and the cyclotron harmonic
energy $\, \om_n\, $ is equal to the scattering resonant energy (see equation [\ref{omres}]).

The reliability and the limits of validity of this approach can be assessed by directly comparing
equation (\ref{absorption}) with the scattering cross sections
(equations [\ref{cross1}] and [\ref{cross2}]), since these expressions are all explicitly written
in terms of a $\delta$-function of the same argument. Figure \ref{Angular} (left panel) shows the ratio
of the absorption to the scattering cross sections for $n=1$ and both ordinary and
extraordinary photons in the range $-2 \leq \log B\leq 2$ and three different directions of the incident
photon (corresponding to $\ct =0,0.5, 1$). The agreement is very good up to $B\sim 0.1$ for both
polarization modes and deviations are within $\sim 20\%$ up to $B\sim 1$.
However, this does not imply that the two descriptions of the photon-electron interaction are to be
regarded as equivalent for $B\la 1$. To claim this, one should prove that they produce also the same angular
redistribution of radiation in the final state. The angular distribution of the scattered photons
is given (in the \ERF) by equation (\ref{npcrossa}) once $B$ and $\theta$ are fixed, while that
of emitted photons by equation (\ref{Transn}). The latter is evaluated in the zero-momentum frame
of the excited electron, exhibits forward-backward symmetry, and lacks any
information on the incoming photon. However, in the case under examination, the electron has
been excited by the absorption of a photon with parallel momentum $\om_n\cos\theta$. In order
to compare the two angular distributions the transition rate must be properly transformed to the
ERF \citep{ah99}, or, alternatively, it should be computed ab initio in the ERF, by exploiting the
general expressions derived in \cite{la86} and valid for an arbitrary value of the electron parallel
momentum. Since $p = \om_n\ct$, this clearly introduces a dependence on the incoming photon kinematical
quantities in the transition rate.

Figure \ref{Angular} (right) shows the angular distribution $\, Q_s(\theta')\, $ of
the scattered photons (upper panel) for $n=1$, a representative value of the incident photon angle,
$\theta=30^\circ$, and different values of the magnetic field, $0.01\leq B\leq 10$.
$\, Q_s\, $ has been computed by normalizing equation (\ref{npcrossa}) to its
maximum value, which occurs at $\theta'=0$ and is the same for any $B$. Photons are more and more
forward-scattered as the field increases. We verified that, as expected, this asymmetry strongly
increases for $\theta\sim 0$ and $\theta\sim 180^\circ$, while the curves become symmetric for
$\theta=90^\circ$. The lower panel shows instead (again, for $n=1$) the fractional difference between
$\, Q_s(\theta')\, $ and the analogous quantity for emission $\, Q_e(\theta')\, $ (here $\theta'\equiv \theta_e$),
obtained normalizing the Lorentz-transformed transition rate (\ref{Transn}).
It is interesting to note that in this case the forward beaming is entirely
due the relativistic Lorentz boosting in going from the \ERFSS\ to the \ERF. As the plot shows, there are
significant differences in the angular distributions even at magnetic
field  below $0.1 {\cal B}_{cr}$. This means that photon-electron scattering can be safely treated as
the superposition of absorption and emission only for resonant photon energies $\la 10$ keV.

Similar results are obtained for transitions involving higher Landau levels, although in this case
sensible differences arise even for lower magnetic field strengths.
However, when $n>1$ major complications arise because of the necessity to discriminate transitions
toward final levels different from the ground state, and to account for the possible changes
of the photon polarization and electron spin orientation.

\section{Scattering probability}
\label{mean}

Let us assume that a photon propagates in a strongly magnetized medium
populated by electrons with a one-dimensional relativistic
velocity distribution along the magnetic field direction

\begin{eqnarray}\label{DistrF}
\su{d n_e}{d\beta} &=& \gamma^{3} n_e f_e(\vec r, \gamma\beta)\, ,
\end{eqnarray}
where $\, f_e = n_e^{-1} d n_e/d(\gamma\beta)\, $  is the
momentum distribution function
and $\, n_e(\vec r) \, $ is the charge number density. As discussed in \S \ref{Spawning}, all
electrons can be taken to be  initially in the ground Landau level. Having in mind the results
obtained in Section \ref{Master}, equation (\ref{TotalCross}), the optical depth after a photon
has travelled an infinitesimal distance $\, d\ell\,$ is
\begin{eqnarray}\label{optdep}
\nonumber
d\tau_{s}  & =&  d\ell\int_{\beta_{min}}^{\beta_{max}} d\beta\,
\su{d n_e}{d\beta} \,(1-\beta\mu)\,\sigma_{s}= \\
& = & d\ell\sum_{n=1}^2 \int_{\beta_{min}}^{\beta_{max}} d\beta\, n_e(\vec r)
\,\gamma^3\, f_e(\vec r, \gamma\beta)\,(1-\beta\mu)\, S_s^n \,
\delta \left[ \gamma (1-\beta\mu) \om^{(L)}-\om_n\right],
\end{eqnarray}
in which $\,[\beta_{min},\beta_{max}]\,$ is the charge velocity spread and
$\,\mu\,$ is the cosine of the propagation angle with respect to the
magnetic field in the stellar frame (LAB). The latter quantity is related
to the same angle as measured in the \ERF\ by the usual transformations
\begin{eqnarray}\label{AngleLorentz}
  \ct &=& \su{\mu-\beta}{1-\beta\mu}\, ,\qquad \st\==\su{\sqrt{1-\mu^2}}{\gamma (1-\beta\mu)}\, .
\end{eqnarray}
The factor $\, (1-\beta\mu)\, $ in the integral (\ref{optdep}) appears because
of the change of reference between \ERF\ and LAB, and, for the sake of clarity, the
argument of the $\delta$-function has been explicitly written in terms of the
energy $ \om^{(L)}$ measured by an observer at rest in the LAB.

The integral (\ref{optdep}) can be readily performed exploiting the well-known
properties of the $\delta$-function. Denoting by $\,\bk\, (k=1,2) $ the two roots
(for each $n$) of the quadratic equation $\gamma (1-\beta\mu) \om^{(L)} -\om_n =0 $,
the $\delta$-function in energy can be transformed into a $\delta$-function in velocity by
\begin{eqnarray}\label{delta}
 \delta [\gamma (1-\beta\mu) \om^{(L)} -\om_n ] &=& \sum_{k=1}^2
\su{\delta(\beta-\bk)}  {\left\vert\gamma^3\, g(\bk)\,\right\vert}\end{eqnarray}
where
\begin{eqnarray}
 g(\beta)\,  &\==& \su{(\mu-\beta\,)\,\om_n }{\sqrt{\gamma^2\,(1-\beta\,\mu)^2+2nB\,(1-\mu^2)}}\,
\label{gprime} \\
 \bk  &\==& \su{1}{m^2+\mu^2}\left[\mu\ \pm \ m\,\sqrt{m^2+\mu^2-1 }\right]\label{betak}\\
 m  &\==& \su{nB}{\om_n } \left\vert 1\--\su{\om_n^2\, (1-\mu^2)}{2nB} \right\vert\, .
\end{eqnarray}
It is worth noticing that with this transformation
the role played  by the resonant photon is replaced by that of a pair of
resonant electrons selected among all charges of the distribution.
Accordingly, the elementary optical depth (\ref{optdep}) for a photon with
initial polarization state $s=(1,2)$, becomes
\begin{eqnarray}\label{optdep1}
d{\tau}_s &=&  {n_e} d\ell\,\sum_{n=1}^2\,\sum_{k=1}^2 \, \su{ (1-\mu\,\beta_k^{(n)})\,
f_e(\vec r, \gamma\bk)\, S_s^{(n)} (B,\theta(\mu,\bk)) }{\vert\, g(\bk)\,\vert}
\=={n_e}\, d\ell\, \sum_{n=1}^2\, A^{(n)}_{s};
\end{eqnarray}
the quantities $\, A^{(n)}_{s}\, $ are implicitly defined  by equation (\ref{optdep1}), and we made
evident the dependence of the resonant factors $S_s^{(n)}$, which are computed in the \ERF,
on the LAB variables.

Once energy, polarization and initial photon direction are fixed, equation
(\ref{optdep1}) can be integrated numerically along the optical path until
a scattering, if any, occurs. As it is apparent from equation
(\ref{gprime}), scattering may occur only when the roots $\bk$ are real,
i.e. only where $\, h(r,\mu) \equiv m^2 +\mu^2-1 \ge 0 $. Since the
function $h$ depends only on position and photon direction, at every point
in the magnetosphere the previous condition discriminates those pairs of
energy and angle for which scattering is possible.

\section{Conclusions}
\label{conc}

Recent models of spectral formation in magnetars called renewed attention on electron-photon
scattering in the presence of ultra-strong magnetic fields. The complete expression for
the QED cross section is known since long \cite[e.g.][]{hd91} but its practical application
in the relativistic regime is numerically challenging. In many astrophysical problems, including
the one which motivated us, it is reasonable to assume that scattering occurs only at
resonance, i.e. when the incident photon frequency equals the cyclotron frequency (or
one of its harmonics). Restricting to resonant scattering introduces a major simplification, and here
we presented explicit expressions for the magnetic Compton cross section in this
particular case. Our main goal has been to provide a complete, workable set of formulae which can be
used in  Monte Carlo simulations of photon scattering in strongly magnetized media.

Our results are fairly general and can be applied to resonant photon scattering under a variety of conditions.
In particular, no assumption is made on the field strength.
Having in mind applications to spectral modelling in
the $\,\sim 0.1$--200 keV range, resonant scattering necessarily occurs where $\, {\cal B}/{\cal B}_{cr}\la 1$.
Under these conditions, and although the expressions we derived are still valid for $\, {\cal B}/{\cal
B}_{cr}>1$, we restricted to the case in which the electron is excited at most up to the second Landau level. We
find that deviations from the non-relativistic limit in both the first and second resonant contributions to the
cross section become significant for $\, {\cal B}/{\cal B}_{cr} \ga 0.1$. The probability that scattering occurs
at the second resonance, which is negligible below $\, {\cal B}/{\cal B}_{cr} \la 0.01$,  becomes sizeable at
higher $B$ and, depending on the scattering angle, it can be up to $\,\sim 30\%\, $ for $\, {\cal B} \sim {\cal
B}_{cr}$. In case the second Landau level is excited, its is more likely that the recoiled electron is left in
the first than in the ground level with the ensuing emission of a new photon (spawning). Using our results for
the cross section together with known expressions for the transition rates, we checked under which conditions
resonant Compton scattering can be treated as the combination of two first-order processes, photon absorption
followed by emission. While the scattering and absorption cross sections differ by at most $\,\sim 20\%\, $  for
$\, {\cal B} \la {\cal B}_{cr}$, the angular distribution of the scattered/emitted photons shows deviations
already at $\,{\cal B} \sim 0.1{\cal B}_{cr}$. Finally, having in mind the 
implementation in a Monte Carlo code  (Nobili, Turolla \& Zane, in 
preparation), we presented an explicit derivation of the scattering 
optical depth along the photon path.

\section*{Acknowledgments}

We are grateful to an anonymous referee whose constructive criticism helpd in improving a previous 
version of this paper. The work of LN and RT is partially supported by INAF-ASI through grant AAE TH-58. 
SZ acknowledges STFC (ex-PPARC) for support through an Advanced Fellowship.

\vfill\eject

\appendix{}\section{}

The expressions below explicitly give the functions $T_\pm^{n\to\ell}(\om,\om',\theta,\theta',B; s, s', f)$
introduced in equation (\ref{FunF}) for the relevant values of $n$ and $\ell$. In the following, to simplify 
notation, only the dependence on $f$ is shown in the argument: $f=0$ refers to the {\it no-flip} and
$f=1$ to the {\it flip} case.
\begin{eqnarray*}\label{AppendA}
 T_+^{1\to0}(0) & = & \su{\om\om'}{(2+\om)^2 }\left[ \su{\PMQ}{\Ome}
                    \,(\V1\ct  - \V3 \st) + \st'\, (\V0\st -\V2\ct)\right] \\
 T_-^{1\to0}(0) & = & \su{1}{(2+\om)^2}\left[\su{\PPQ\om'\st'}{2B\Ome}\,(2B\,\V2+\V0 \om^2 \st\ct)
 + (2B \V1 + \V3 \om^2\st\ct)\right]\\
 T_+^{2\to0}(0) & = & \su{\om^2\om'^2\st\st'}{ 2B\,(2+\om)^2}\left[\su{\PMQ}{\Ome}
        \,(\V1\ct -\V3\st) + \st'\, (\V0\st -\V2\ct)\right] \\
 T_-^{2\to0}(0)  & = & \su{\om\om'\st\st'}{2 B\,(2+\om)^2}
\left[\su{\PPQ\om'\st'}{4B\Ome}\, (4B\,\V2 + \V0\om^2\st\ct)+(4B\,\V1+\V3\om^2\st\ct)\right]\\
T_+^{2\to1}(0) & = & \su{\sqrt{2\,}\,\om^2\om' \st\, }{4B^{3/2}(2+\om)^2}
\left\{\su{\PMQ}{\Ome}\, (2B-\om'^2\stq') (\V1\ct -\V3\st) \++ \su{}{}\right.\\
&& + \left.\st'\left[\su{ 2B\, (2+\om-2\om')}{\Ome}-\om'^2\stq'\right](\V0\st -\V2\ct)\right\}\\
 T_-^{2\to1}(0) & = & \su{\sqrt{2\,}\,\om\st\,}{8\, B^{3/2}\,(2+\om)^2}\left\{\su{\PPQ\om'\st'
 (4B-\om'^2\stq')}{2B\Ome}\,(4B\V2+\om^2 \V0 \st\ct) + \phantom{\su{|}{}}\right.\\
  && \left. + (4B-2\om'^2\stq')\, (4B\V1+\V3\om^2\st\ct) +
     \su{\om'^2\stq' (2+\om)}{\Ome}\, (4B\V5+\V4\om^2\st\ct)\right\}\\
T_+^{2\to1}(1) & = & \su{\om^2\st}{2B\,\Omega\, (2+\om)^2}
\left\{ \su{}{} \PPQ \om'\st'\, (\V2\ct -\V0\st)+(2+\om)(2B-\om'^2\stq')(\V1\ct -\V3\st) \++ \right.\\
&&  \left. \su{}{} + \Omega\om'^2\stq'(\V5\ct -\V4\st) \right\}\\
T_-^{2\to1}(1) & = & -\, \su{\om\,\om'^2\st\st'}{8\,\Omega\, B^{2}\,(2+\om)^2}\,
\left\{\PMQ\om' \st'\,(4B\, \V5 + \V4\om^2\ct\st) +\right. \\
&& 
\left.+\left[4B -\om'\stq' (2+\om)\right]\, (4B\,\V2+\V0\om^2\st\ct)\su{}{} \right\} \\
\end{eqnarray*}
where
\begin{eqnarray*}\label{AppendB}
 \PMQ &=&  (2+\om)\ct'- \om\ct\quad , \qquad \PPQ =\om (4+2\om-\om')\ct-\om' (2+\om) \ct' \quad , \qquad \Ome=2+
\om-\om'
\end{eqnarray*}
and
\begin{eqnarray*}\label{AppendC}
\qquad\V0 \== &\Mat{\st\st'}{0}{0}{0}     \, \, , \, \, \, \,   & \qquad \V1
\==\Mat{\ct\ct'}{\ct}{-\ct'}{1}\\
\qquad\V2 \== &\Mat{-\ct\st'}{0}{\st'}{0}  \, \, , \,  \, \, \, & \qquad \V3
\==
\Mat{-\st\ct'}{-\st}{0}{0} \\
\qquad\V4 \== &\Mat{-\st\ct'}{\st}{0}{0}    \, \, , \, \, \, \, & \qquad \V5
\== \Mat{\ct\ct'}{-\ct}{-\ct'}{-1}.
\end{eqnarray*}
In each expression the functional form for given values of the photon polarization, $s, \,s'$, is recovered inserting the matrix
element $\V{k} (s,s')$ .

\label{lastpage}
\end{document}